\input psfig
\documentclass{aa}
\begin{document}
\thesaurus{13(13.07.1; 03.19.2)}
\title{Ballerina -- Pirouettes in Search of Gamma Bursts}

   \author{	S. Brandt 		\inst{1}	\and
		N. Lund 		\inst{1}	\and
		H. Pedersen		\inst{2}	\and
		J. Hjorth		\inst{2}
		for the Ballerina Collaboration\thanks{
		The Ballerina Collaboration presently consist of members from:
		Danish Space Research Institute;
             	Astronomical Observatory, University of Copenhagen;
		Institute of Physics and Astronomy, Aarhus University;
		Aalborg University;
		Stockholm Observatory;
		Helsinki Observatory;
		University of Birmingham;
		Max Planck f{\"u}r Extraterrestische Physik, Garching;
		LAEFF, INTA, Madrid; 
		Astronomiches Institut Potsdam
		}	
          }

   \offprints{N. Lund}

   \institute{	Danish Space Research Institute,
              	Juliane Maries Vej 30, DK-2100 Copenhagen {\O}, Denmark
         \and
             	Astronomical Observatory, University of Copenhagen,
		Juliane Maries Vej 30, DK-2100 Copenhagen {\O}, Denmark
}

\date{Received ; accepted }

\maketitle

\begin{abstract}
The cosmological origin of gamma ray bursts has now been 
established with reasonable certainty.  
Many more bursts will need to be studied to establish the typical 
distance scale, and to map out the large diversity in properties 
which have been 
indicated by the first handful of events. We are proposing Ballerina, 
a small satellite to provide accurate positions and new data on the 
gamma-ray bursts. We anticipate a detection rate an order of magnitude 
larger than obtained from Beppo--SAX.

\end{abstract}


\section{Introduction}
The crucial observation of an X-ray transient following 
a cosmic gamma ray burst (GRB) was made by the Italian-Dutch satellite 
Beppo--SAX (Costa et al. \cite{Costa}). Several GRBs
with X-ray and optical afterglows have subsequently been found.
In several cases, it has been possible to identify a remote galaxy as a 
likely host to the burst source, and to derive its distance. 
The huge luminosities, and the very short time scales involved in GRBs, 
may possibly be explained by merging, or collapsing, compact objects 
such as neutron stars or black holes (e.g. Paczy{\'n}ski \cite{Pacz}). 
However, 
the observed energy spectra of the bursts, and the details of the 
energy transport and conversion, are still to be explained. 
Models involving relativistic fireballs and jets have been proposed 
(e.g. Dar \cite{Dar}).
To date no theory has been able to explain all of the observed GRB 
phenomena, and it has become a challenge for modern astrophysics.

\section{The Ballerina Mission}
It remains an objective to ensure the earliest 
detection of the X-ray afterglow. Ballerina will for the 
first time allow systematic studies of the soft X-ray emission in the 
time interval from only a few minutes after the onset of the burst 
to a few hours later. Ballerina will, on its own provide observations 
in an uncharted region of parameter space.  
Positions of GRB sources with accuracy better than $1\arcmin$ will be 
distributed within a few minutes of the burst.

Secondary objectives of the Ballerina mission includes observations
of the earliest phases of the outbursts of X-ray novae and other X-ray
transients. 

In addition to the autonomous observations of events detected on-board, 
Ballerina may on short notice be commanded from the ground to 
execute observations on objects identified by other observatories.  
\begin{figure}
\psfig{file=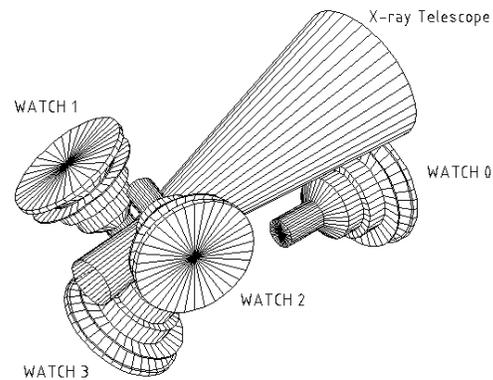,width=8.2cm}
\caption[]
{
A possible accommodation of the Ballerina payload with four WATCH
wide field monitors and a grazing incidence X-ray 
telescope
}
\label{balfig}
\end{figure}

Ballerina is a spacecraft in the 100 kg class.
The payload consists of an all-sky monitor and a grazing incidence 
X-ray telescope. 
A compact accommodation of these key elements is shown in 
Fig.~\ref{balfig}.
Ballerina is one of four missions currently under study for the 
Danish Small Satellite Programme. 
If selected, Ballerina will be launched in 2002. 

\subsection{The All-Sky Monitor}
The all-sky monitor consists of four Rotation Modulation Collimator 
instruments, arranged in a 
tetrahedron configuration to cover the full sky. Each unit is similar 
to the WATCH instruments flown successfully on the 
{EURECA} and GRANAT missions (Lund \cite{Lund85}; 
Castro-Tirado \cite{AJCT}; Sazonov et al. \cite{Sazonov}; 
Brandt \cite{Brandt}). 
The detector used is a mosaic of NaI and CsI scintillators, with 
an energy range of 6--120 keV.
The established threshold sensitivity of WATCH for a GRB of 5 
second duration is $10^{-5}$ $\mbox{erg/cm}^2$. 
Bursts of this fluence can be located to better than $1\degr $ 
diameter ($2 \sigma$). 

Placed in a high (Molniya-type) orbit with an efficiency of 55\%,
Ballerina will observe $\approx 80 $ 
bursts per year distributed over the full sky. 
About 10 of these will be too 
close to the Sun to be observed with the X-ray telescope, 
but the remaining 70 bursts
will be located to better than 0.5 $\arcmin$.

The capacity of the telemetry link for Ballerina will be 
significantly better
than available on GRANAT and on EURECA. 
We do not expect the telemetry to
be a limiting factor for transmitting the data from the observed GRBs.

Efficient rejection of false triggers will be a high priority 
objective for the on-board software. 
The decision to slew the satellite to a new pointing
will only be taken when the existence of a GRB (or X-ray transient)
source has been confirmed by localizing the source consistently using 
two independent datasets. 
This is very efficient in rejecting false triggers. However,
the real problem will be to manage the computational effort to 
search for
two consistent localizations in the many data set combinations 
possible from a marginal trigger.

Satellite attitude slews will be executed solely by controlling 
the speed of each of the four WATCH modulators. 
Magneto-torquers will be used for momentum dumping. 

\subsection{The Grazing Incidence X-Ray Telescope}
This instrument will be a smaller version of the instrument used on 
ROSAT with 
a focal length of 60 cm equipped with a CCD focal plane detector. 
Our design is based on a telescope 
with nested mirror shells, 
the required effective area being 50 $\mbox{cm}^2$. The field 
of view is $1.5\degr$, and the mirror shape will be optimized 
for achieving a rms resolution $< 30\arcsec$ over the entire field. 
The mirror fabrication technique is similar to the one, which has been 
used with great success for the mirrors for Beppo--SAX and 
for ESA's XMM mission. 
The energy response will be in the range 0.5--2 keV.  

The sensitivity of the Ballerina telescope will not be much different 
from the sensitivity of the Beppo--SAX telescopes. 
Consequently, it should be possible to follow the 
afterglows for 24--48 hours. In fact, the bursts
detected by the Ballerina all-sky monitor will on average be brighter, 
than the bursts seen by the SAX Wide Field Cameras. 
Therefore, we expect that also the afterglows will be brighter, 
and may be followed for a longer time. The orbit of
Ballerina will permit up to 7 hours of uninterrupted observations of
an afterglow source. This will allow to detect possible
deviations from the simple power law decay of the afterglow.

\subsection{Burst Operations Sequence}
The satellite normally operates in a three-axis stabilized mode, 
performing survey observations and follow-up on previously detected 
bursts, waiting for a new burst to occur. 
The GRB is detected by one of the four wide-field cameras. 
An initial position with accuracy better than $1\degr$ is derived, and 
transmitted to the ground.
A slew is initiated to acquire the afterglow with the pointed 
X-ray telescope. 	
The satellite will be able to slew to the new target 
in 20--70 s, depending on the distance of the slew. 
A fine burst position (better than $30\arcsec$) is then determined.

The observations of the decaying afterglow are automatically scheduled 
to take spacecraft constraints into consideration. 
The full sky is accessible, except a cone around the Sun of 
$45\degr $ half angle. Real time communication (at a low data rate) 
is continuously available. Source position
information will be downlinked and distributed to the community 
as soon as it is available.

Modifications to the observation plan may also be uploaded from the 
ground.

\section{Conclusion}
Cosmic gamma ray bursts is a continuing enigma for astrophysics. 
We are proposing Ballerina, a small satellite to provide accurate GRB 
positions at a rate an order of magnitude larger than Beppo--SAX.
The Ballerina spectral data will place powerful constraints on 
the theoretical work being done to understand GRBs. 
Establishing the cosmological distances to GRBs, and 
understanding their nature will provide us with a new and 
independent probe of the structure of the Universe -- 
complementary to supernovae, quasars and clusters of galaxies.

\begin{acknowledgements}
Part of this work was supported by the Danish Space Advisory Board
\end{acknowledgements}

\end{document}